\documentclass[preprints,article,accept,moreauthors,pdftex]{Definitions/mdpi}
\firstpage{1} 
\pubvolume{1}
\issuenum{1}
\articlenumber{0}
\pubyear{2022}
\copyrightyear{2022}
\externaleditor{Academic Editor: {Firstname Lastname} 
}
\datereceived{} 
\dateaccepted{} 
\datepublished{} 
\hreflink{https://doi.org/} 

\Title{RetroComposer: Composing Templates for Template-Based Retrosynthesis Prediction}

\TitleCitation{RetroComposer: Composing Templates for Template$-$Based Retrosynthesis Prediction}

\Author{{Chaochao Yan} 
 $^{1}$, Peilin Zhao $^{2,}$*, Chan Lu $^{2}$, Yang Yu $^{2}$ and Junzhou Huang $^{1,}$*}

\AuthorNames{Chaochao Yan, Peilin Zhao, Chan Lu, Yang Yu and Junzhou Huang}

\AuthorCitation{Yan, C.; Zhao, P.; Lu, C.; Yu, Y.; Huang, J.}

\address{%
$^{1}$ \quad Computer Science and Engineering, University of Texas at Arlington, {Arlington, TX 76019, USA;} 
 chaochao.yan@mavs.uta.edu (C.Y.); jzhuang@uta.edu (J.H.)\\
$^{2}$ \quad Tencent AI Lab, {Shenzhen 518054, China};  masonzhao@tencent.com (P.Z.); sherryllu@tencent.com (C.L.); kevinyyu@tencent.com (Y.Y.)}

\corres{Correspondence: masonzhao@tencent.com, jzhuang@uta.edu}

\abstract{
The main target of retrosynthesis is to recursively decompose desired molecules into available building blocks.
Existing template$-$based retrosynthesis methods follow a template selection stereotype and suffer from  limited training templates, which prevents them from discovering novel reactions.
To overcome this limitation, we propose an innovative  retrosynthesis prediction framework that can compose novel templates beyond training templates.
As far as we know, this is the first method that uses machine learning to compose reaction templates for retrosynthesis prediction.
Besides, we propose an effective reactant candidate scoring model that can capture atom$-$level transformations, which helps our method outperform previous methods on the USPTO$-$50K dataset. 
Experimental results show that our method can produce novel templates for 15 USPTO$-$50K test reactions that are not covered by training templates.
We have released our source implementation.
}


\begin{document}

\section{Introduction}

Retrosynthesis plays a significant role in organic synthesis planning, in~which target molecules are  recursively decomposed into available commercial building blocks.
This analysis mode was firstly formulated in the pioneering work~\cite{corey1969computer, corey1991logic} and now is one of the fundamental paradigms in modern chemical society.
Since then numerous retrosynthesis prediction algorithms have been proposed to aid or even automate the retrosynthesis analysis.
However, the~performance of existing methods is still not satisfactory. 
The massive search space is one of the major challenges of retrosynthesis considering that on the order of $10^7$ compounds and reactions~\cite{gothard2012rewiring} have been reported in synthetic--organic knowledge.
The other challenge is that there are often multiple viable retrosynthesis pathways and it is challenging to decide the most appropriate route since the feasibility of a route is often compounded by several factors, such as reaction conditions, reaction yield, potential toxic byproducts, and~the availability of potential reactants~\cite{yan2020retroxpert}.

Most of existing machine$-$learning$-$powered retrosynthesis methods focus on the single-step version. These methods are broadly grouped into template$-$based and template$-$free major categories. 
Templates$-$free methods~\cite{liu2017retrosynthetic, zheng2019predicting, yan2020retroxpert, shi2020graph, sacha@megan2021, sun2021towards} usually rely on deep learning models to directly generate reactants.
One effective strategy is to formulate the retrosynthesis prediction as a sequence translation task and~generate SMILES~\cite{weininger1988smiles} sequences directly using sequence$-$to$-$sequence models such as Seq2Seq~\cite{liu2017retrosynthetic}, SCROP~\cite{zheng2019predicting}, and~AT~\cite{tetko2020state}. 
SCROP~\cite{zheng2019predicting} proposes to use a second transformer to correct the initial wrong predictions. Translation$-$based methods are simple and effective, but~lack interpretability behind the prediction.
Another representative paradigm is to first find a reaction center and split the target accordingly to obtain hypothetical units called synthons, and~then generate reactants incrementally from these synthons, such as in RetroXpert~\cite{yan2020retroxpert}, G2Gs~\cite{shi2020graph}, RetroPrime~\cite{wang2021retroprime}, and~GraphRetro~\cite{somnath2021learning}.

On the other hand, template$-$based methods are receiving less attention than the rapid surge of template$-$free methods. Template$-$based methods conduct retrosynthesis based on either hand$-$encoded rules~\cite{szymkuc2016computer} or automatically extracted retrosynthesis templates~\cite{coley2017computer}.
Templates encode the minimal reaction transformation patterns,= and~are straightforwardly interpretable.
The key procedure is to select applicable templates to apply to targets~\cite{coley2017computer, segler2017neural, segler2018planning, dai2019retrosynthesis}. 
Template$-$based methods have been criticized for the limitation that they can only infer reactions covered by training templates and cannot discover novel reactions~\cite{segler2017modelling, yan2020retroxpert}.

In this work, we propose a novel template$-$based single$-$step retrosynthesis framework to overcome the mentioned limitation. 
Unlike previous methods only selecting from training templates, we propose to compose templates with basic template building blocks (molecule subgraphs) extracted from training templates.
Specifically, our method composes templates by first selecting appropriate product and reactant molecule subgraphs iteratively, and~then annotates atom  transformations between the selected subgraphs.
This strategy enables our method to discover novel templates from training subgraphs, since the reaction space of our method is the exponential combination of these extracted template subgraphs.
What is more, we design an effective reactant scoring model that can capture atom$-$level transformation information.
Thanks to the scoring model, our method achieves state$-$of$-$the$-$art (SOTA) Top$-$1 accuracy of 54.5\% and 65.9\% on the USPTO$-$50K dataset without and with reaction types, respectively. Our contributions are summarized as:
(1) we propose a first$-$ever template$-$based retrosynthesis framework to compose templates, which can discover novel reactions beyond the training data;
(2) we design an effective reactant scoring model that can capture atom$-$level transformations, which contributes significantly to the superiority of our method;
(3) the proposed method achieves 54.5\% and 65.9\% Top$-$1 accuracy on the benchmark  dataset USPTO$-$50K without and with reaction types, respectively, which establishes the new SOTA~performance.
The code is available at \url{https://github.com/uta-smile/RetroComposer} {(accessed on 15 September 2022).} 

\section{Related~Work}
Recently there has been an increasing amount of work using machine learning methods to solve the retrosynthesis problem.
These methods can be categorized into template$-$based~\cite{coley2017computer,  segler2017neural, segler2018planning, baylon2019enhancing, dai2019retrosynthesis} and template$-$free approaches~\cite{liu2017retrosynthetic, shi2020graph, yan2020retroxpert, somnath2021learning, tu2021permutation}. 
Template$-$based methods extract templates from training data and build models to learn the corresponding relationship between products and templates. 
RetroSim~\cite{coley2017computer} selects the templates based on the fingerprint similarity between products and reactions.
NeuralSym~\cite{segler2017neural} uses a neural classification model to select corresponding templates.
However, this method does not scale well with an increasing number of templates.
To mitigate the problem,~\cite{baylon2019enhancing} adopts a multi$-$scale classification model to select templates according to a manually defined template hierarchy.
GLN~\cite{dai2019retrosynthesis} proposes a graph logic network to model the decomposed template hierarchy by first selecting reaction centers within the targets, and then only consider templates that contain the selected reaction centers.
The decomposition strategy can reduce the search space significantly. 
GLN models the relationship between reactants and templates jointly by applying selected templates to obtain reactants, which are also used to optimize the model~simultaneously.

Template$-$free methods do not rely on retrosynthesis templates. Instead, they construct models to predict reactants from products directly.
Translation$-$based methods~\cite{zheng2019predicting, tetko2020state, irwin2021chemformer, mao2021molecular} use SMILES to represent molecules and treat the problem as a sequence$-$to$-$sequence task.
MEGAN~\cite{sacha@megan2021} treats the retrosynthesis problem as a graph transformation task, and~trains the model to predict a sequence of graph edits that can transform the product into the reactants. 
To imitate a chemist's approach to the retrosynthesis,  two$-$step methods~\cite{shi2020graph, yan2020retroxpert, wang2021retroprime, somnath2021learning} first perform reaction center recognition to obtain synthons by disconnecting targets according to the reaction center, and~then generate reactants from the synthons.  
G2Gs~\cite{shi2020graph} treats the reactant generation process as a series of graph editing operations and utilizes a variational graph generation model to implement the generation process. 
RetroXpert~\cite{yan2020retroxpert} converts the synthon into SMILES to generate reactants as a translation task.
GraphRetro~\cite{somnath2021learning} also adopts a similar framework and generates the reactants by attaching leaving groups to synthons.
Dual model~\cite{sun2021towards} proposes a general energy$-$based model framework that integrates both sequence$-$ and graph$-$based models, and~performs consistent training over forward and backward prediction~directions.

\section{Preliminary~Knowledge}
\unskip

\subsection{Retrosynthesis and~Template}

\begin{figure}[H]
	\includegraphics[width=0.8\textwidth]{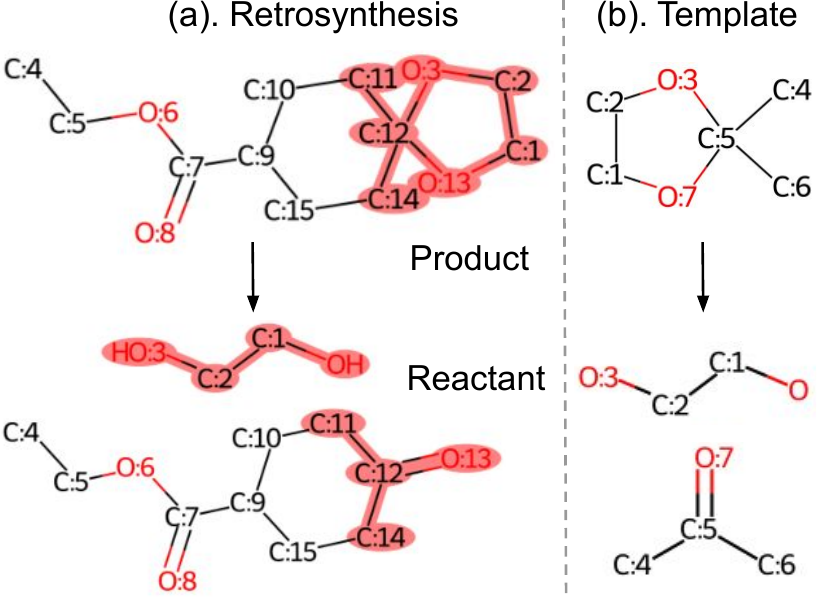}
	\caption{A retrosynthesis example from USPTO$-$50K dataset and its template extracted using an open$-$source toolkit.  Note that the product and reactant are atom$-$mapped. The~product and reactant subgraphs in (\textbf{b}) are highlighted in pink within the product and reactant molecule graphs in (\textbf{a}), respectively.}
	\label{fig:reaction_template_example}
\end{figure}

Single$-$step retrosynthesis predicts a set of reactant molecules given a target product, as shown in Figure~\ref{fig:reaction_template_example}a. 
Note that the product and reactant molecules are atom$-$mapped, which ensures that every product atom is uniquely mapped to a reactant atom.
Templates are reaction rules extracted from chemical reactions. They are composed by reaction centers and encode the atom and bond transformations during the reaction process.  
The illustrated template in Figure~\ref{fig:reaction_template_example}b consists of a product subgraph (upper) and reactant subgraphs (lower). The~subgraph patterns are highlighted in pink within the corresponding molecule~graphs.

\unskip

\subsection{Molecule Graph~Representation}

The graph representation of a molecule or subgraph pattern is denoted as $G(\mathcal{V}, \mathcal{E})$, where $ \mathcal{V} $ and  $ \mathcal{E} $ are the set of graph nodes (atoms) and edges (bonds), respectively. Following previous work~\cite{dai2019retrosynthesis, yan2020retroxpert}, each bond is represented as two directed edges. 
Initial node and edge features can be easily collected for the learning~purpose.

\subsection{Graph Attention~Networks} \label{section:gat}

Graph neural networks~\cite{gilmer2017neural} are especially good at learning node$-$ and graph$-$level embeddings of molecule data. 
In this work, we adapt graph attention networks (GATs) \cite{velickovic2018graph} to incorporate bond features.
The GAT layer updates a node embedding by aggregating its neighbor's information.
The modified GAT concatenates edge embeddings with the associated incoming node embeddings before each graph message passing. 
The input of the GAT layer is node embeddings $ \{ v_i | \forall i \in \mathcal{V} \} $ and edge features $ \{ e_{i, j} | (i, j) \in \mathcal{E} \} $, and~the output updated node embeddings $ \{ v_{i}' | \forall i \in \mathcal{V} \} $. Each node embedding is updated with a shared parametric function $t_{\theta}$:\vspace{12pt} 
\begin{equation}
v_i' = t_{\theta}(v_i, \mathrm{AGGREGATE}(\{ [v_j || e_{i, j}] |\forall j \in \mathcal{N}(i)\})),
\end{equation}
where $ \mathcal{N}(i) $ are neighbor nodes of $v_i$ and $||$ is the concatenation operation. The~$\mathrm{AGGREGATE}$ of GAT adopts an attention$-$based mechanism to adaptively weight the neighbor information. A~scoring function $c(i, j)$ computes the importance of the neighbor node $j$ to node~$i$:
\begin{equation}
c(i, j) = \mathrm{LeakyReLU}(w^T  [ \boldsymbol{W}_1  v_i || \boldsymbol{W}_1 v_j || \boldsymbol{W}_2 e_{i, j}] ),
\end{equation}
where $w$ is a learnable vector parameter and each $\boldsymbol{W}$ is a learnable matrix parameter. These importance scores are normalized using the Softmax function across the neighbor nodes $\mathcal{N}(i)$ of the node $i$ to obtain attention weights:
\begin{equation}
\alpha(i, j) = \mathrm{Softmax}_j(c(i, j) ) = \frac{\exp (c(i, j)) }{\sum_{j' \in \mathcal{N}(i) }{ \exp(c(i, j'))}}.
\end{equation}

The modified GAT instances $t_{\theta}$  and updates the node embedding as the non$-$linear function $\sigma$ activated weighted$-$sum of the transformed embeddings of its neighbor nodes:
\begin{equation}
v_i' = \sigma ( \sum_{j \in \mathcal{N}(i)} { \alpha(i, j) * \boldsymbol{W}_3 [\boldsymbol{W}_1  v_j || \boldsymbol{W}_2 e_{i, j} ]}).
\end{equation}

GAT is usually stacked by multiple layers and enhanced with multi$-$head attention~\cite{vaswani2017attention}. Please refer to~\cite{velickovic2018graph} for more~details.

\subsection{Graph$-$Level~Embedding}
After obtaining the output node embeddings from the GAT, a~graph $\mathrm{READOUT}$ operation can be used to obtain the graph$-$level embedding.
Inspired by~\cite{xu2018representation}, we aggregate and concatenate the output node embeddings from all GAT layers to learn structure$-$aware node representations from different neighborhood ranges:
\begin{equation} \label{eq:graph_readout}
    \mathrm{emb}_G = \mathrm{READOUT}( \{v_{i, 1} || v_{i, 2} || ... || v_{i, L} | \forall i \in \mathcal{V} \}).
\end{equation}
where $v_{i, l}$ is the output embedding of node $i$ after the $l$th GAT layer. 
The $\mathrm{READOUT}$ can be any permutation$-$invariant operation (e.g., $\mathrm{mean, sum, max}$). We adopt the global soft attention layer from~\cite{li2015gated} as the $\mathrm{READOUT}$ function for molecule graphs due to its excellent~performance.


\section{Methods}

We propose to compose retrosynthesis templates from a predefined set of template building blocks; then, these composed templates are applied to target products  to obtain the associated reactants.
Unlike previous template$-$based methods~\cite{coley2017computer, segler2017neural, segler2018planning, dai2019retrosynthesis} only selecting from training templates, our method can discover novel templates that are beyond the training templates.
To further improve the retrosynthesis prediction performance, we design a scoring model to evaluate the suitability of product and candidate reactant pairs. 
The scoring procedure acts as a verification step and~it plays a significant role in our~method.

\begin{figure}[H]

\begin{adjustwidth}{-\extralength}{0cm}
\centering 
\includegraphics[width=1.3\textwidth]{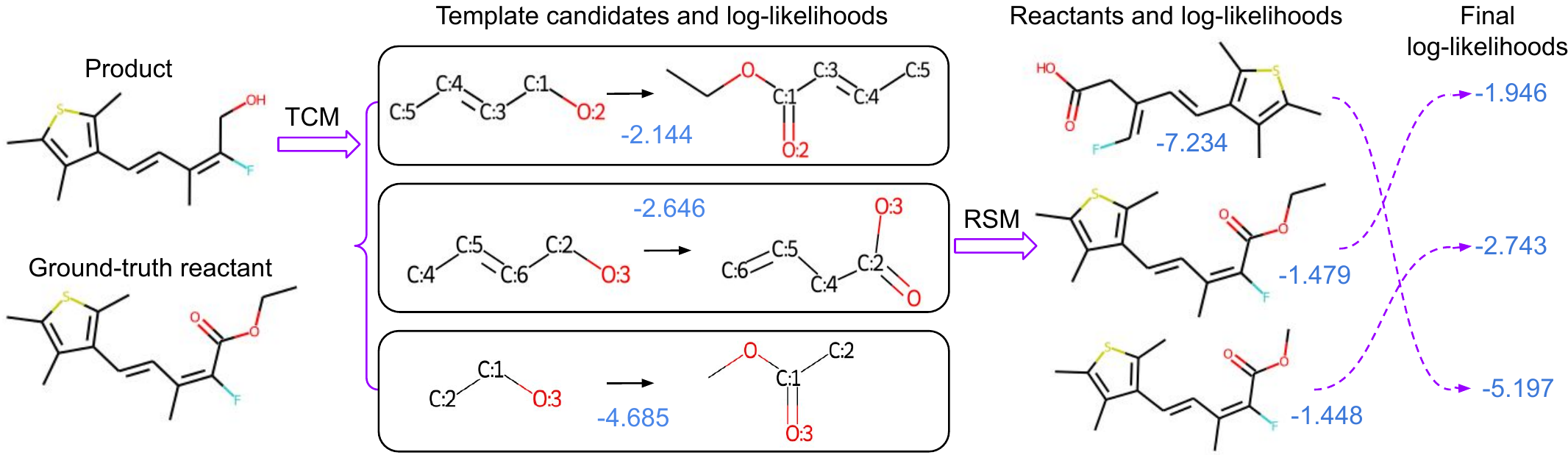}
\end{adjustwidth}
\caption{The overall pipeline of our proposed method. Given the desired product as shown at the top left, single$-$step retrosynthesis finds the ground$-$truth reactant as shown at the bottom left.
Numbers indicated in blue are the corresponding log$-$likelihoods of our models, and~the log$-$likelihoods of the template composer model (TCM) and the reactant scoring model (RSM) are combined to obtain the final ranking of the reactants. In~this example, combining log$-$likelihoods of TCM and RSM helps to find the correct Top$-$1 {reactant.} 
}
\label{fig:pipeline}
\end{figure}

The overall pipeline of our method is shown in Figure~\ref{fig:pipeline}. 
Our method tackles retrosynthesis in two stages.
The first stage is to compose retrosynthesis templates with a TCM, which composes retrosynthesis templates by selecting  template building blocks and then assembling them.
In the second stage, the~obtained templates are applied to the target product to generate the associated reactants.
After that, we utilize a powerful RSM to evaluate the generated reactants for each product. 
During evaluation, the~probability scores of both stages are linearly combined to rank Top$-$K reactant predictions. 
In following sections, we will detail each stage of our~method.

\subsection{Compose Retrosynthesis~Templates}

\begin{figure}[H]
\centering
\includegraphics[width=0.99\textwidth]{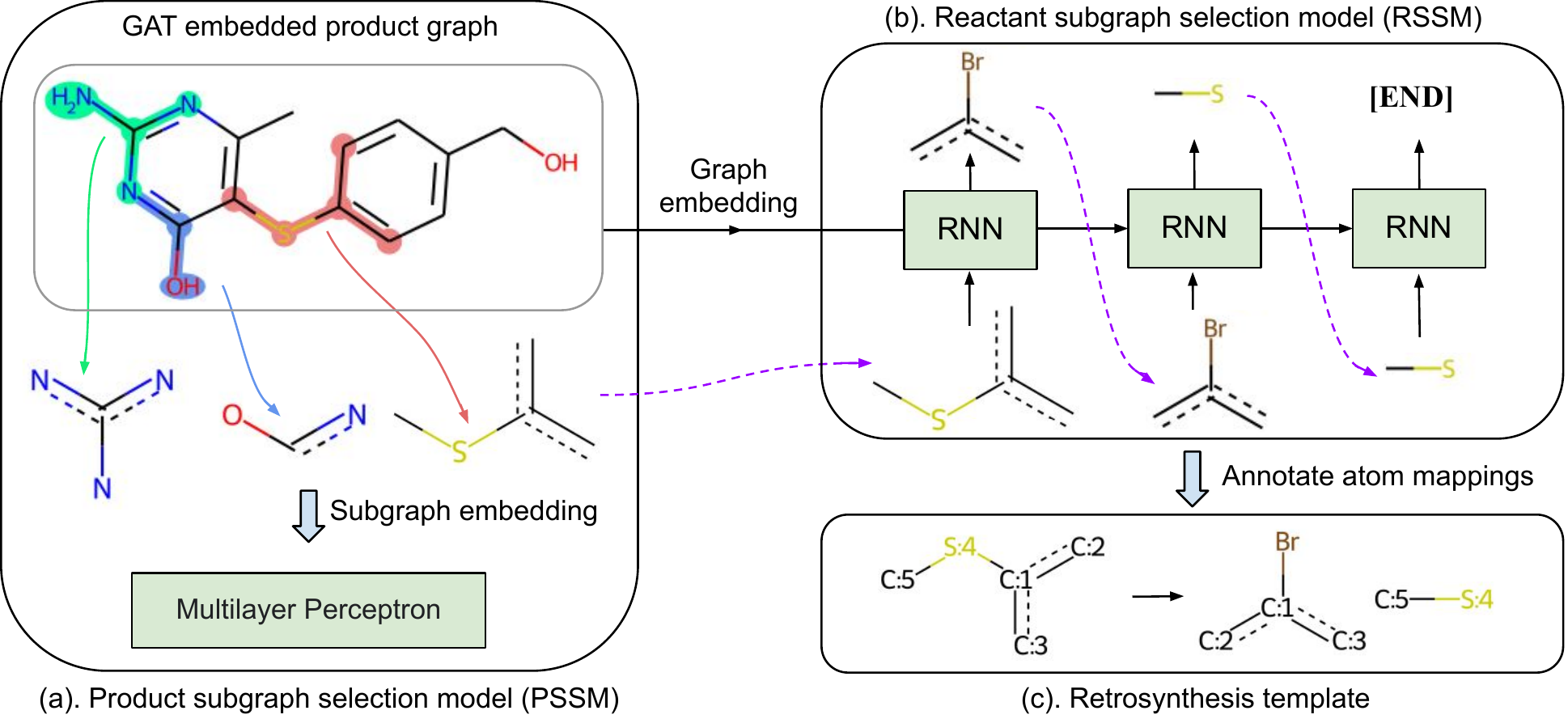}
\caption{The workflow of our template composer model: (\textbf{a}) selecting a proper product subgraph from product subgraph candidates with PSSM, (\textbf{b}) selecting reactant subgraphs sequentially from reactant subgraph vocabulary with RSSM, and~(\textbf{c}) annotating atom mappings between the product and reactant subgraphs to obtain a template.
}
\label{fig:overview}
\end{figure}

Template$-$based retrosynthesis methods are criticized for their limitation of not generalizing to unseen reactions, since all existing template$-$based methods follow a similar procedure to select applicable templates from the extracted training templates. 
To overcome the above limitation, we propose a different pipeline to find template candidates.
As illustrated in Figure~\ref{fig:overview}, our method first selects product and reactant subgraphs sequentially from the corresponding subgraph vocabularies, which is detailed in Section~\ref{section:subgraph_selection}.
Then, these selected subgraphs are assembled into templates with properly assigned atom mappings, as detailed in Section~\ref{section:annote_atom_mappings}. 
As far as we know, this is the first attempt to compose retrosynthesis templates instead of simple template selection.
During evaluation, a beam search algorithm~\cite{tillmann2003word} is utilized to find Top$-$K predicted templates.
Reactants can be obtained by applying templates to the target molecule.

\unskip

\subsubsection{Subgraph~Selection} \label{section:subgraph_selection}
We denote a subgraph pattern as $f$, the~product and reactant subgraphs for a template as $f_p$ and $f_r$, respectively, and~the product and reactant subgraph vocabulary for the dataset as $\mathcal{F}_P$ and $\mathcal{F}_R$, respectively.
To build the product subgraph vocabulary $\mathcal{F}_P$ and reactant subgraph vocabulary $\mathcal{F}_R$, retrosynthesis templates extracted from training data are split into separate subgraphs to collect unique subgraph patterns. 
We build separate vocabularies for the product and reactant subgraphs due to their essential difference.
Product subgraphs represent reaction centers and are more generalizable, while reactant subgraphs may contain extra leaving groups, which are more specific to the reaction type and the desired target.
We find this strategy works well in~practice.

\subsubsection{Product Subgraph~Selection} \label{section:product_subgraph}
To compose retrosynthesis templates for a desired target, the~first step is to choose proper $f_p$ from the vocabulary $\mathcal{F}_P$. In~this work, we focus on the single$-$product reactions; therefore, there is only a single product subgraph pattern. 
Note that there may be multiple viable retrosynthesis templates for each reaction, so each target may have several applicable product subgraphs. The~set of applicable product subgraphs are denoted as $\mathcal{F}_a$. 
Starting with any applicable product subgraph in $\mathcal{F}_a$ may obtain a applicable retrosynthesis template for the target. Here, $\mathcal{F}_a \subseteq  \mathcal{F}_P$ because all applicable product subgraphs must be in the vocabulary $\mathcal{F}_P$.

Each product molecule graph $G_p$ contains only a limited set of  candidate subgraphs $\mathcal{F}_c$ predefined in the vocabulary $\mathcal{F}_P$. Three candidate subgraphs  are illustrated in Figure~\ref{fig:overview}a.
The candidate subgraphs for each target can be obtained offline by checking the existence of every product subgraph from $\mathcal{F}_P$ in the product graph $G_p$. 
Therefore, we only need to consider the candidate subgraphs $\mathcal{F}_c$ to guide the selection process~\cite{dai2019retrosynthesis} when selecting a product subgraph.
Here, $\mathcal{F}_a \subseteq \mathcal{F}_c \subseteq \mathcal{F}_P$ since the candidate subgraphs $\mathcal{F}_c$ must contain all applicable~subgraphs.

In this situation, the~product subgraph selection can be regarded as a multi$-$label classification problem and starting with any applicable product subgraph in $\mathcal{F}_a$ can obtain a viable retrosynthesis template.
However, it is not optimal to train the product subgraph selection model with binary cross$-$entropy loss (BCE) as in the multi$-$label classification setting, since it predicts the applicability score independently for each  $f \in \mathcal{F}_c $ without considering their interrelationship. 
Note that the absolute applicability scores of subgraphs in $\mathcal{F}_c $ do not matter here; what really matters is the ranking of these applicability scores, since the beam search is adopted to find a series of template candidates during model inference. 
While a Softmax classifier can consider the relationship of all subgraphs in $\mathcal{F}_c $, it cannot be directly applied to PSSM, since it is not suitable for the multi$-$label case.
Inspired by Softmax, we propose a novel negative log$-$likelihood loss for the PSSM:
\begin{equation}\label{eq:pssm_loss}
L_{\mathrm{PSSM}} =  \log\frac{  \arg\min_{f \in \mathcal{F}_a} o_f } { {\arg\min_{f \in \mathcal{F}_a} o_f }  + \sum_{f \in \mathcal{F}_c \setminus \mathcal{F}_a} o_f},
\end{equation}
where $o_f$ is the exponential of PSSM output logits for subgraphs in $\mathcal{F}$, $|\mathcal{F}|$ is the size of $\mathcal{F}$, and~$\setminus$ is set subtraction. 
In the above loss function, the~numerator is the minimal exponential output for all applicable subgraphs in $\mathcal{F}_a$, which is considered as the ground$-$truth class proxy in the Softmax classifier. 
The extra item in denominator is the summation of exponential output of all inapplicable subgraphs in $\mathcal{F}_c$. 
The intuition is that we always optimize the PSSM to increase the prediction probability for the least probable applicable subgraph, so the model is driven to generate large scores for all applicable subgraphs $\mathcal{F}_c$ while considering interrelationships of candidate subgraphs. The~novel loss outperforms BCE loss in our experiments. Detailed experimental comparison results between the proposed loss function Equation~(\ref{eq:pssm_loss}) and BCE loss can be found in the  {experiment section.}

PSSM  scores candidate subgraphs $\mathcal{F}_c $ based on their subgraph embeddings. 
As shown in  Figure~\ref{fig:overview}a,
to obtain subgraph embeddings, the~nodes of product molecule graph $G_p$ are first encoded with the modified GAT that is detailed in Section~\ref{section:gat}. 
The embedding $\mathrm{emb}_f$ of the subgraph $f$ is gathered as the average embedding of subgraph $f-$associated nodes in $G_p$, and~then these embeddings are fed into a multilayer perceptron (MLP) for subgraph selection.
Here, for a subgraph $f$, the~$\mathrm{READOUT}$ function is implemented as the arithmetic average for its simplicity and efficiency.
Note that this is different from GLN~\cite{dai2019retrosynthesis}, in which product graphs and candidate subgraphs are considered as separate graphs and embedded independently.
Our strategy to reuse node embeddings is more efficient and can learn more informative subgraph embedding since the neighboring structure of a subgraph is also incorporated during the message passing procedure of GAT.
Besides, our method can naturally handle multiple equivalent subgraph situations in which the same subgraph appears multiple times within the product~graph.

\subsubsection{Reactant Subgraph~Selection}

The second step of the subgraph selection is to choose reactant subgraphs $f_r$ from the vocabulary $\mathcal{F}_R$, which is ordered according to the subgraph frequency in training data, so that $f_r$ is also determinedly ordered. 
With minor notation abuse, $f_r$ also denotes an ordered sequence of reactant subgraphs in the following~content.

Since the number of reactant subgraphs is undetermined, we build the reactant subgraph selection model based on the recurrent neural network (RNN), as illustrated in Figure~\ref{fig:overview}b, and~formulate reactant subgraph selection as the sequence generation.
The hidden state of RNN is initialized from the product graph embedding
$\mathrm{emb}_{G_p}$ as defined in Equation~(\ref{eq:graph_readout}) to explicitly consider the target product, and~the start token is the product subgraph $f_p$ selected in the previous procedure (Section \ref{section:product_subgraph}). Furthermore, an extra end token $ [ \mathrm{END} ]$ is appended to reactant subgraph sequence $f_r$. At~each time step, the~RNN output is fed into a MLP for the token classification. 
For the start token $f_p$, we reuse product subgraph embeddings obtained previously (Section \ref{section:product_subgraph}) since we find it provides better performance than embedding the token in the traditional one$-$hot embedding~manner.

\subsubsection{Annotate Atom~Mappings} \label{section:annote_atom_mappings}

Given $f_p$ and $f_r$, the~final step is to annotate the atom mappings between $f_p$ and $f_r$ to obtain the retrosynthesis template, as shown in Figure~\ref{fig:overview}c.
A subgraph pattern $f$ can also be represented in the SMARTS string, and~we use open source toolkit Indigo's (\url{https://github.com/epam/Indigo} {(accessed on 20 March 2022)}
) $\mathrm{automap}()$ function to build atom mappings.
We empirically find about 70\% of USPTO$-$50K  training templates can be successfully annotated with correct atom mappings. 
To remedy this deficiency, we keep a memo of training templates and associated $f_p$ and $f_r$. 
During evaluation, the~predicted $f_p$ and $f_r$ are processed with $\mathrm{automap}()$ if not found in the memo. 

\subsection{Score Predicted~Reactants}

After a retrosynthesis template is composed, reactants can be easily obtained by applying the template to the target using   $\mathrm{RunReactants}$ from RDKit~\cite{landrum2006rdkit} or the $\mathrm{run\_reaction}()$ function from RDChiral~\cite{coley2019rdchiral}.
To achieve superior retrosynthesis prediction performance, it is important to verify that the predicted reactants can generate the target successfully.
The verification is achieved by scoring the reactants and target pair, which is formulated as a multi$-$class classification task where the true reactant set is the ground$-$truth~class.

To serve the verification purpose, we build a reactant scoring model based on the modified GAT. 
Product molecule graph $G_p$ and reactant molecule graph $G_r$ are first input into a GAT to learn atom embeddings. 
Since the target and generated reactants are atom$-$mapped as in Figure~\ref{fig:reaction_template_example}a, for~each atom in $G_p$, we can easily find its associated atom in $G_r$. 
Inspired by WLDN~\cite{jin2017predicting}, we define a fusion function $ \mathrm{F}(n_a^p, n_{a'}^r)$ to combine embeddings of a product atom $a$ and its associated reactant atom $a'$:
\begin{equation} \label{eq:fusion}
\mathrm{F}(n_a^p, n_{a'}^r) = \boldsymbol{W}_4  (n_a^p - n_{a'}^r) ||  \boldsymbol{W}_5 (n_a^p + n_{a'}^r),
\end{equation}
where $||$ indicates the concatenation operation and $\boldsymbol{W}$ is a matrix that halves the node embedding dimension so that the concatenated embedding restores the original~dimension.

The fused atom embeddings are regarded as new atom features of $G_p$, which are input into another GAT to learn the graph$-$level embedding $emb_G$.
In this way, the~critical difference between the product and reactant can be better captured since our RSM can incorporate higher order interactions between fused atom embeddings through the message passing process of GAT. Previous retrosynthesis methods score reactants by modeling the compatibility of reactant and product at the molecule level without considering the atom$-$level~embedding.

The graph$-$level embedding $emb_G$ is then fed into a simple MLP composed of two fully$-$connected layers to output a compatibility score. The~final probability score is obtained by applying a Softmax function to the compatibility scores of all candidate reactants associated to the~target. 

Our scoring model is advantageous since it operates on atom$-$level embeddings and is sensitive to local transformations between the product and reactants, while the existing method GLN~\cite{dai2019retrosynthesis} takes only molecule$-$level representations as the input. Therefore, GLN cannot capture atom$-$level transformations and has a weaker distinguishing~ability.

The log$-$likelihoods of our TCM and RSM model predictions are denoted as  $l_{TCM} = \log(\mathcal{P}(\mathcal{T} | P)) $  and $l_{RSM} = \log(\mathcal{P}(R | P)) $, respectively.
The predicted reactants are finally ranked according to the linear combination value of $ \lambda * l_{TCM} + (1 - \lambda ) * l_{RSM}, 0 \leq \lambda \leq 1 $.
The formulation can be understood as:
\begin{equation}  \label{eq:combine_score}
\begin{split}
 & \lambda * \log(\mathcal{P}(\mathcal{T} | P)) + (1 - \lambda ) * \log(\mathcal{P}(R | P)) \\
 = & \log(\mathcal{P}(\mathcal{T} | P)^\lambda * \mathcal{P}(R | P)^{1 - \lambda}),
\end{split}
\end{equation}
where $\mathcal{P}(\mathcal{T} | P)$ is the probability of that the template $\mathcal{T}$ is applicable to the given product $P$ and $\mathcal{P}(R | P)$ is the probability of the reactant set $R$ for the given product $P$.
When combined together,  $\mathcal{P}(\mathcal{T} | P) * \mathcal{P}(R | P)$ approximates the joint probability distribution $\mathcal{P}(\mathcal{T}, R | P)$.
Hyper$-$parameter $ \lambda $ regulates the relative importance of $\mathcal{P}(\mathcal{T} | P)$ and $\mathcal{P}(R | P)$.
The optimal $ \lambda $ can be determined by the~validation.




\section{Experiment and~Results}
\unskip

\subsection{Dataset and~Preprocessing}
Our method is evaluated on the standard benchmark dataset USPTO$-$50K~\cite{schneider2016what} under two settings (with or without reaction types) to demonstrate its effectiveness. 
USPTO$-$50K is derived from USPTO granted patents~\cite{lowe2012extraction} and is composed of 50,000 
reactions annotated with 10 reaction types.
More detailed dataset information can be found in the appendix \ref{appendix:datainfo}.
We split reaction data into training/validation/test sets at an 8:1:1 ratio, in the same way as previous work~\cite{coley2017computer, dai2019retrosynthesis}.
Since the original annotated mapping numbers in the USPTO dataset may result in unexpected information leakage (\url{https://github.com/uta-smile/RetroXpert} {(accessed on 20 March 2022)}), 
 we first preprocess the USPTO reactions to re$-$assign product mapping numbers according to the canonical atom order, as suggested by RetroXpert~\cite{yan2020retroxpert}.
The atom and bond features are similar to the previous work~\cite{yan2020retroxpert} and~reaction types are converted into one$-$hot vectors concatenated with the original atom features.


Following RetroXpert~\cite{yan2020retroxpert}, we extract templates from training reactions using RDChiral~\cite{coley2019rdchiral}. We can obtain 10386 unique templates in total for the USPTO$-$50K training data and~94.08\% of test reactions are covered by these training templates.
The gathered templates are split into product and reactant subgraphs, from which mapping numbers are further removed  to obtain the subgraph vocabularies $\mathcal{F}_P$ of size 7766 and $\mathcal{F}_R$ of size~4391. 

For each target molecule, we find its candidate subgraphs $\mathcal{F}_c$ using graph matching algorithms and applicable templates by checking if the ground$-$truth reactant can be obtained when each training template is applied to the target. 
The applicable subgraphs $\mathcal{F}_a$ then can be obtained easily from the acquired applicable templates. 
Since the exact graph matching process might be time$-$consuming, we extract the fingerprint for each molecule/sub$-$molecule to filter those impossible subgraphs.
For the subgraph screening purpose, we adopt the $\mathrm{PatternFingerprint}$ from RDKit and use a fingerprint size of~1024.

\subsection{Evaluation}

Following previous methods~\cite{dai2019retrosynthesis, yan2020retroxpert}, we use beam search~\cite{tillmann2003word} to find Top$-$50 template predictions during evaluation, which are applied to targets to collect candidate reactants.
The collected reactants and targets are the experimental data for RSM.
The predicted reactants are finally ranked according to the combined log$-$likelihood of TCM and RSM. 
The evaluation metric for retrosynthesis prediction is the Top$-$K exact match accuracy, which is the percentage of reactions where the ground truth reactant set is within the top K predictions.

\subsection{Implementation}

Our model is implemented using PyTorch~\cite{paszke2019pytorch} and PyTorch Geometric~\cite{fey2019fast}. The~adapted GAT model is built based on the source implementation of Pretrain$-$GNN~\cite{hu2019strategies}.
The TCM model is composed of a modified GAT and a simple RNN model.
The embedding dimension is set as 300 for all embeddings for simplicity.
The number of GAT layers is six. We adopt GRU~\cite{cho2014properties} as the RNN implementation in  TCM; the~number of GRU layers is two and both its embedding and hidden size are 300. 
We add a self$-$loop to each graph node following~\cite{dai2019retrosynthesis, yan2020retroxpert}. 
We use the parametric rectified linear unit (PReLU)  \cite{he2015delving} comprehensively as the activation function in our model.
We replace the original batch normalization~\cite{ioffe2015batch} layer with a layer normalization~\cite{ba2016layer} layer after each GAT layer, since we find layer normalization provides better performance in our experiments.
We adopt Equation~(\ref{eq:graph_readout}) as the graph $\mathrm{READOUT}$ operation.
A simple MLP is applied to product subgraph embeddings to select the proper product subgraph.
The MLP is composed of two linear layers, and the PReLU activation function is placed between the two linear layers.
We also use a dropout~\cite{srivastava2014dropout} layer with a dropout rate of 0.3 in the~MLP.

The RSM model is composed of two GATs and a MLP head, and~the GAT uses the same settings as in the TCM except that each GAT is composed of three layers.
Product and reactant graphs are embedded with the first GAT model.
Note that for reactions with multiple reactants, we regard the disconnected molecule graphs as a single large graph.
Once the fused atom embeddings are obtained, the~new product molecule graphs with fused atom embeddings are input into the second GAT.
The composition of the MLP head is similar to that in TCM. The~RSM model is also trained in multi$-$process mode for~acceleration.

Both TCM and RSM are optimized with the Adam~\cite{kingma2014adam} optimizer with default settings, and~the initial learning rates are 0.0003 and 0.00005 for  TCM and RSM, respectively.
The learning rate is adjusted with the $\mathrm{CosineAnnealingLR}$ scheduler during training.
The models are trained in multi$-$process mode on a single GTX 1080 Ti GPU for acceleration.
 TCM is trained with batch size 32;
it only takes about two hours to train  TCM for 80 epochs.
 RSM training takes about 6 hours for 20 epochs.
The final model parameters are saved and loaded later for inference.
We repeat all experiments three times and report the mean performance as default. 
We find our model is quite robust to the hyper$-$parameters, and~most of the model settings are copied from~\cite{hu2019strategies} as they are given. 
We slightly tune the model hyper$-$parameters, such as learning rate and batch size, manually on validation data to achieve the best~results.

\subsection{Main~Results}
We decide the optimal value of $\lambda$  according to validation performance.
Specifically, we set $\lambda$ as 0.4 for both experimental settings (with/without reaction types).
We use these optimal settings in all experiments unless explicitly stated.
Detailed ablation study about $\lambda$ are included in  \ref{section:ablation_lambda}.


\subsubsection{Retrosynthesis Prediction~Performance}

We compare our RetroComposer with existing methods on the standard benchmark dataset USPTO$-$50K, and~report comparison results in Table~\ref{tab:retro-comparison-with-sota}.
The results of RetroXpert have been updated by the authors (\url{https://github.com/uta-smile/RetroXpert} {(accessed on 20 March 2022)}). 
For both evaluation settings (with or without reaction types), our method outperforms previous methods by a significant margin in seven out of eight compared Top$-$K~metrics.
\begin{table}[H]
    \caption{Retrosynthesis evaluation results (\%) on USPTO$-$50K. Existing methods are grouped into two categories. Our method RetroComposer belongs to the template$-$based methods. The~best results in each column are highlighted in bold. RetroXpert* results have been updated by the authors in their GitHub repository (\url{https://github.com/uta-smile/RetroXpert} {(accessed on 20 March 2022)}).}
\label{tab:retro-comparison-with-sota}
    \newcolumntype{C}{>{\centering\arraybackslash}X}
\begin{tabularx}{\textwidth}{cCCCCCCCC}
    \toprule
     \multirow{2}{*}{\vspace{-4pt} \textbf{Methods}} & \multicolumn{4}{c}{{\textbf{Without Reaction Types}}} & \multicolumn{4}{c}{{\textbf{With Reaction Types}}}  \\ \cmidrule{2-9}
    & \textbf{Top$-$1}& \textbf{Top$-$3} & \textbf{Top$-$5} & \textbf{Top$-$10} & \textbf{Top$-$1} & \textbf{Top$-$3} & \textbf{Top$-$5} & \textbf{Top$-$10} \\
    \midrule
    \multicolumn{9}{c}{Template$-$free methods} \\ \midrule
    SCROP~\cite{zheng2019predicting} & 43.7 & 60.0 & 65.2 & 68.7 & 59.0 & 74.8 & 78.1 & 81.1 \\ 
    G2Gs~\cite{shi2020graph} & 48.9 & 67.6 & 72.5 & 75.5 & 61.0 & 81.3 & {86.0} & {88.7} \\
    MEGAN~\cite{sacha@megan2021} & 48.1 & 70.7 & 78.4 & 86.1 & 60.7 & 82.0 & 87.5 & \textbf{91.6} \\
    RetroXpert* \cite{yan2020retroxpert} & 50.4 & 61.1 & 62.3 & 63.4 & 62.1 & 75.8 & 78.5 & 80.9  \\
    RetroPrime~\cite{wang2021retroprime} & 51.4 & 70.8 & 74.0 & 76.1 & 64.8 & 81.6 & 85.0 & 86.9 \\
    AT~\cite{tetko2020state} & 53.5 & $-$ & 81.0 &  85.7 & $-$ & $-$ & $-$ & $-$ \\
    GraphRetro~\cite{somnath2021learning} & {53.7} & 68.3 & 72.2 & 75.5 &  63.9 & 81.5 & 85.2 & 88.1  \\ 
    Dual model~\cite{sun2021towards} & 53.6 & {70.7} & {74.6} & {77.0} & {65.7} & {81.9} & {84.7} & {85.9} \\
     \midrule
     \multicolumn{9}{c}{Template$-$based methods} \\ \midrule
    RetroSim~\cite{coley2017computer} & 37.3 & 54.7 & 63.3 & 74.1 & 52.9 & 73.8 & 81.2 & 88.1 \\ 
    NeuralSym~\cite{segler2017neural} & 44.4 & 65.3 & 72.4 & 78.9 & 55.3 & 76.0 & 81.4 & 85.1 \\
    GLN~\cite{dai2019retrosynthesis} & 52.5 & 69.0 & 75.6 & 83.7 & 64.2 & 79.1 & 85.2 & 90.0 \\
    Ours & \textbf{54.5} & \textbf{77.2} & \textbf{83.2} & \textbf{87.7} & \textbf{65.9} & \textbf{85.8} & \textbf{89.5} & {91.5} \\
    TCM only & {49.6} & {71.7} & {80.8} & {86.4} & {60.9} & {82.3} & {87.5} & {90.9} \\
    RSM only & {51.8} & {75.7} & {82.4} & {87.3} & {64.3} & {84.8} & {88.9} & {91.4} \\
    \bottomrule
    \end{tabularx}

\end{table}

Specially,  our RetroComposer achieves 54.5\% Top$-$1 accuracy without reaction types, which improves on the previous best template$-$based method GLN~\cite{dai2019retrosynthesis} significantly by 2.0\%
and also outperforms existing SOTA template$-$free methods Dual model and GraphRetro.
Besides, our method achieves 77.2\% Top$-$3 accuracy, which improves on the Top$-$3 accuracy 70.8\% of RetroPrime~\cite{wang2021retroprime} by 6.4\%, and~87.7\% Top$-$10 accuracy, which improves on the Top$-$10 accuracy 85.7\% of AT~\cite{tetko2020state} by 2.0\%.

When reaction types are given, our method also obtains the best Top$-$1 accuracy, 65.9\%, among all methods and outperforms GLN by 1.7\%.
Compared with template$-$free methods GraphRetro and Dual model, our method outperforms the SOTA Dual model (65.7\%) by 0.2\% and GraphRetro significantly by 2.0\% in Top$-$1 accuracy.
As for the Top$-$10 accuracy, our method achieves 91.5\%, which is slightly lower than 91.6\% of MEGAN~\cite{sacha@megan2021}.

As the ablation study, we report results with only TCM or RSM.
With only either TCM or RSM, the~model performance is largely degraded. Without~reaction types,  TCM only achieves 49.6\% Top$-$1 accuracy while RSM achieves only 51.8\%.
With reaction types, TCM only achieves 60.9\% Top$-$1 accuracy while RSM achieves only 64.3\%.
Since TCM and RSM score retrosynthesis from different perspectives and are complementary, their results can be combined to achieve the best performance.
Particularly, our method achieves  54.5\% and 65.9\% Top$-$1 accuracy when combining TCM and RSM according to Equation~(\ref{eq:combine_score}).

The superior performance demonstrates the effectiveness of our method. Particularly, the~superiority of our method is more significant in real world applications where reaction types are unknown. 
What is more, our Top$-$10 accuracy is already quite high. 
This indicates that our method can usually find the best reactant set for the target in a few candidates.
This is especially important for multi$-$step retrosynthesis scenarios, in~which the number of predicted reaction paths may grow exponentially with the retrosynthesis path~length.

\subsubsection{Ablation Study of PSSM~Loss}
\label{section:ablation_pssm_loss}
We experimentally show that our proposed loss function Equation~(\ref{eq:pssm_loss}) for PSSM outperforms the BCE loss.
For all ablation experiments, we find the optimal value of hyper$-$parameter $\lambda$ independently and report the best results for a fair comparison.
The comprehensive experimental results are reported in Table~\ref{tab:loss-function}. 

Without given reaction types, our method with Equation~(\ref{eq:pssm_loss}) as PSSM loss achieves the best Top$-$1 and Top$-$3 accuracy results, outperforming the BCE loss in Top$-$1 and Top$-$3 accuracy by 1.4\% and 1.5\%, respectively.
With known reaction types, our method with Equation~(\ref{eq:pssm_loss}) as PSSM loss outperforms BCE loss by 0.6\% in Top$-$1 accuracy. While BCE loss can achieve better Top$-$5 and Top$-$10 results in both settings, our proposed loss function Equation~(\ref{eq:pssm_loss}) can achieve better Top$-$1 accuracy.
The retrosynthesis prediction emphasizes more Top$-$1 accuracy,; therefore, we adopt Equation~(\ref{eq:pssm_loss}) as the PSSM loss in our~method.

For all experiments, combining the TCM and RSM scores can always achieve the best performance, which proves the effectiveness of our~strategy.

\begin{table}[H]
    \caption{Ablation study results (\%) of two {different PSSM loss} 
 functions: our proposed Equation~(\ref{eq:pssm_loss}) and BCE. The bold indicates the best results. \label{tab:loss-function}}
    \newcolumntype{C}{>{\centering\arraybackslash}X}
\begin{tabularx}{\textwidth}{CCCCCCC}
    \toprule
    \textbf{Types} & \boldmath{$L_{\mathrm{PSSM}}$} & \textbf{Methods}  & \textbf{Top$-$1} & \textbf{Top$-$3} & \textbf{Top$-$5} & \textbf{Top$-$10}  \\ \midrule
    \multirow{6}{*} { Without} &  \multirow{3}{*}{Equation (\ref{eq:pssm_loss})} & Ours & {\textbf{54.5}} 
 &  \textbf{77.2} & 83.2 & 87.7  \\
    & & TCM only  & {49.6} & {71.7} & {80.8} & {86.4} \\
    & & RSM only & {51.8} & {75.7} & {82.4} & {87.3} \\
    \cmidrule(l){2-7}
    & \multirow{3}{*}{BCE} &  Ours  & 53.1 & 77.1 & \textbf{83.8} & \textbf{89.2} \\
    & & TCM only  & 46.5 & 69.9 & 78.5 & 86.9 \\
    & & RSM only  & 51.2 & 75.7 & 82.9 & 88.6 \\
    \midrule
    \multirow{6}{*} { With} &  \multirow{3}{*}{Equation (\ref{eq:pssm_loss})} & Ours  & \textbf{65.9} & 85.8 & 89.5 & 91.5  \\
    & & TCM only & {60.9} & {82.3} & {87.5} & {90.9} \\
    & & RSM only & {64.3} & {84.8} & {88.9} & {91.4} \\
    \cmidrule(l){2-7}
    &  \multirow{3}{*}{BCE}  & Ours & 65.3 & \textbf{85.9} & \textbf{90.3} & \textbf{92.6}  \\
    & & TCM only & 58.5 & 81.8 & 87.6 & 91.5 \\
    & & RSM only & 64.2 & 85.4 & 89.6 & 92.4 \\
    \bottomrule
    \end{tabularx}
\end{table}
\unskip

\subsubsection{Ablation Study of Hyper$-$Parameter $\lambda$} \label{section:ablation_lambda}
We conduct the ablation study of $\lambda$ and report results in Table~\ref{tab:hyper-lambda}; when $\lambda = 0.4$, the best Top$-$1 accuracy is achieved for both settings.
Note that with only RSM ($ \lambda = 0$), the~Top$-$1 accuracy 64.3\% already outperforms the previous best template$-$based method GLN of 63.2\% \cite{dai2019retrosynthesis} with given  reaction types.
This demonstrates the effectiveness of our RSM.
With only TCM ($ \lambda = 1.0$), the~performance has an appreciable gap with the existing methods.
In our method, each generated set of subgraphs may have multiple associated templates due to the uncertainty of product subgraphs and atom transformations.
Therefore, there may be multiple top$-$tier predictions that cannot be distinguished with only TCM. With~a little help from RSM ($ \lambda = 0.9$), these top$-$tier predictions can be differentiated and the Top$-$1 accuracy significantly~boosted.


The $l_{RSM}$ indicates the likelihood of retrosynthesis templates, while $l_{TCM}$ scores each reaction by looking at the detailed atom transformations. These two terms are complementary and combined together to achieve the best~performance.

\begin{table}[H]
    \caption{Top$-$1 accuracy (\%) with different $ \lambda $ values. The bold indicates the best results.\label{tab:hyper-lambda}}
    \newcolumntype{C}{>{\centering\arraybackslash}X}
\begin{tabularx}{\textwidth}{lCCCCCCCCCCC}
    \toprule
    \boldmath{$\lambda$} & \textbf{0} & \textbf{0.1} & \textbf{0.2} & \textbf{0.3} & \textbf{0.4} & \textbf{0.5} & \textbf{0.6} & \textbf{0.7} & \textbf{0.8} & \textbf{0.9} & \textbf{1.0}  \\ \midrule
    Without types  & 51.8 & 53.3 & 53.9 & {\textbf{54.5}} 
 & \textbf{54.5} & 54.4 & 54.1 & 53.6 & 53.0 & 52.3 & 49.6 \\
    With types  & 64.3 & 65.2 & 65.6 & 65.7 & \textbf{65.9} & \textbf{65.9} & 65.6 & 65.1 & 64.7 & 64.4 & 60.9 \\ 
    \bottomrule
    \end{tabularx}
\end{table}
\unskip

\subsubsection{Novel~Templates}
\begin{figure}[H] 
	\includegraphics[width=0.9\textwidth]{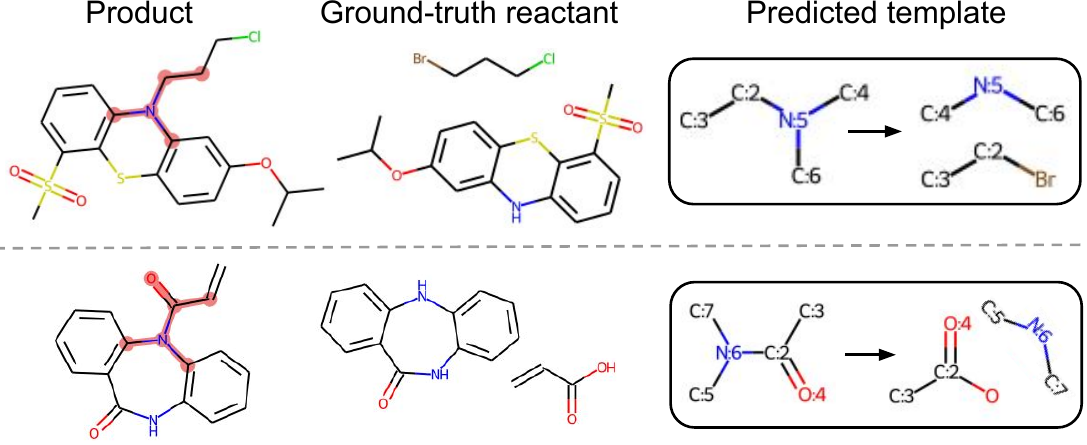}
	\caption{Our method successfully finds valid templates for two test reactions that are not covered by training data. The~matched product subgraphs are highlighted in pink for better~visualization.}
	\label{fig:novel_reactions}
\end{figure}

Different from existing methods, our method can find novels templates that are not in training data. 
Our model predicts different templates based on different possible reaction centers for a given target. 
For example, an~amide formation template and alkylation template may both be applied in the same target molecule, and~our model can predict suitable templates very well and give reasonable corresponding reactants for such cases.
For the 5.92\% of test reactions that are not covered by training templates, our algorithm can predict relevant templates very well for most reaction types, although~it fails in some heterocyclic formation reactions.
This is because there are very few reaction data on such reactions in USPTO$-$50K.
Particularly, our method successfully discovers chemically valid templates for 15 uncovered test reactions, which confirms that our method can find novel reactions.
Two such examples are illustrated in Figure~\ref{fig:novel_reactions}. 


\unskip

\section{Discussion and~Conclusions}
In this work, we propose a novel template$-$based retrosynthesis prediction framework that composes templates by selecting and assembling molecule subgraphs.
Besides, experimental results confirm that the proposed strategy can discover novel reactions. Although~currently our method can find only a few novel templates, we believe our method can inspire the community to explore further in this direction to improve models' ability to find more novel reactions.
To further improve the ranking accuracy, we present a novel reactant scoring model to rank candidate reactants by taking into account atom$-$level transformations.
Our method significantly outperforms previous methods and sets new SOTA performance on the USPTO$-$50K, which proves the effectiveness of our~method.

{We tried to adapt our method to run on the USPTO$-$full dataset~\cite{lowe2012extraction}, but~find it needs non$-$trivial effort to manually handle edge cases  due to noisy reactions (such as wrong mapping numbers) from USPTO$-$full, since our methods rely on correct mapping numbers to extract templates as well as build the reactant scoring model.
We have released our source implementation and encourage the community to help adapt our method to the USPTO$-$full dataset.}

\vspace{6pt} 



\authorcontributions{Conceptualization, C.Y. and P.Z.; methodology, C.Y. and P.Z.; formal analysis, Y.Y.; investigation, C.L.; writing—original draft preparation, C.Y. and C.L.; writing—review and editing, C.Y. and P.Z.; supervision, J.H.; project administration, J.H.; funding acquisition, J.H. All authors have read and agreed to the published version of the~manuscript.}

\funding{This work was partially supported by US National Science Foundation IIS$-$1553687 and Cancer Prevention and Research Institute of Texas (CPRIT) award (RP190107).}
\institutionalreview{{Not applicable.} 
}

\informedconsent{{Not applicable.} 
}


\dataavailability{The experimental dataset USPTO$-$50K can be downloaded at \url{http://pubs.acs.org/doi/suppl/10.1021/acs.jcim.6b00564/suppl_file/ci6b00564_si_002.zip} {(accessed on 20 March 2022).}
}

\conflictsofinterest{The authors declare no conflict of interest. The~funders had no role in the design of the study; in the collection, analyses, or~interpretation of data; in the writing of the manuscript, or~in the decision to publish the~results.}




\appendixtitles{no} 
\appendixstart
\appendix
\section[\appendixname~\thesection]{}\appendixtitles{yes}
\subsection[\appendixname~\thesubsection]{USPTO$-$50K Dataset Information}\label{appendix:datainfo}

{The} 
 USPTO$-$50K consists of 50,000 reactions that are annotated with 10 reaction types; the detailed distribution of reaction types is displayed in the below Table~\ref{tab:reaction-type-distribution}.
The imbalanced reaction type distribution makes the retrosynthesis prediction more~challenging.

\begin{table}[H]
\caption{Distribution of 10 recognized reaction~{types.} 
}
    \label{tab:reaction-type-distribution}  
    \newcolumntype{C}{>{\centering\arraybackslash}X}
\begin{tabularx}{\textwidth}{Clr}
    \toprule
     \textbf{Type} & \textbf{Reaction Type Name} & \textbf{Number of Reactions} \\
    \midrule
    1& Heteroatom alkylation and arylation & 15,204 \\
    2& Acylation and related processes & 11,972 \\
    3& C$-$C bond formation & 5667 \\
    4& Heterocycle formation& 909 \\
    5& Protections& 672 \\
    6& Deprotections & 8405 \\
    7& Reductions& 4642 \\
    8& Oxidations& 822 \\
    9& Functional group interconversion & 1858 \\
    10& Functional group addition (FGA)& 231 \\
    \bottomrule
    \end{tabularx}
\end{table}

We can extract 10,386 unique templates from the training data, and~94.08\% of test reactions are covered by these templates. For~each product molecule, there are an average of 35.19 candidate subgraphs, which are denoted as $\mathcal{F}_c $ in Section~\ref{section:product_subgraph}.
Among these subgraphs, there are an average of 2.02 applicable subgraphs denoted as $\mathcal{F}_a$ for each~target.

\begin{table}[H]
\caption{Statistical results of templates and~reactions. \# is the short for "number".}
    \label{tab:templates_statistics}  
    \newcolumntype{C}{>{\centering\arraybackslash}X}
\begin{tabularx}{\textwidth}{Cr}
    \toprule
    \# total templates &  10,386 \\
    \# unique product subgraphs  &  7766 \\
    \# unique reactant subgraphs  &  4391 \\
    Test reactions coverage by training templates & 94.08\% \\
    Average \# contained product subgraphs per mol &  35.19 \\
    Average \# applicable product subgraphs per mol &  2.02 \\
    Average \# templates per reaction &  2.23 \\
    Average \# reactants per reaction &  1.71 \\
    \bottomrule
    \end{tabularx}
\end{table}
\unskip

\subsection[\appendixname~\thesubsection]{Atom and Bond Features}\label{appendix:atom-bond-features}

Following~\cite{yan2020retroxpert}, we use similar bond and atom features to build molecule graphs as listed in Tables~\ref{tab:bond-feature} and \ref{tab:atom-feature}. These features can be easily extracted using the chemistry toolkit~RDKit.

\begin{table}[H]
\caption{Bond features used in our method. These features are one$-$hot~encoding.}
\label{tab:bond-feature}
    \newcolumntype{C}{>{\centering\arraybackslash}X}
\begin{tabularx}{\textwidth}{CcC}
    \toprule
    \textbf{Feature} & \textbf{Description} & \textbf{Size} \\
    \midrule
    Bond type &  Single, double, triple, or~aromatic. & 4 \\
    Conjugation &  Whether the bond is conjugated. & 1 \\
    In ring & Whether the bond is part of a ring. & 1 \\
    Stereo & None, any, E/Z or cis/trans. & 6 \\
    \bottomrule
    \end{tabularx}
\end{table}
\unskip

\begin{table}[H]
\caption{Atom features used in our method. All features are one$-$hot encoding, except~the atomic mass is a real number scaled to be on the same order of magnitude. The~reaction type is applicable for type conditional~setting.}
    \label{tab:atom-feature}
    \newcolumntype{C}{>{\centering\arraybackslash}X}
\begin{tabularx}{\textwidth}{Ccr}
    \toprule
    \textbf{Feature} & \textbf{Description} & \textbf{Size} \\
    \midrule
    Atom type &  Type of atom (ex. C, N, O), by~atomic number. & 17 \\
    \# Bonds &  Number of bonds the atom is involved in. & 6 \\
    Formal charge & Integer electronic charge assigned to atom. & 5 \\
    Chirality & Unspecified, tetrahedral CW/CCW, or~other. & 4 \\
    \# Hs & Number of bonded Hydrogen atom. & 5 \\
    Hybridization &  sp, sp2, sp3, sp3d, or~sp3d2. & 5 \\
    Aromaticity & Whether this atom is part of an aromatic system. & 1 \\
    Atomic mass & Mass of the atom, divided by 100.  & 1 \\
    \midrule
    Reaction type & The specified reaction type if it exists.  & 10 \\
    \bottomrule
    \end{tabularx}
\end{table}

\begin{adjustwidth}{-\extralength}{0cm}

\reftitle{References}

\end{adjustwidth}
\end{document}